\documentclass[twoside,twocolumn,9pt]{article}
\usepackage{extsizes}
\usepackage[super,sort&compress,comma]{natbib} 
\usepackage[version=3]{mhchem}
\usepackage[left=1.5cm, right=1.5cm, top=1.785cm, bottom=2.0cm]{geometry}
\usepackage{balance}
\usepackage{mathptmx}
\usepackage{sectsty}
\usepackage{graphicx} 
\usepackage{lastpage}
\usepackage[format=plain,justification=justified,singlelinecheck=false,font={stretch=1.125,small,sf},labelfont=bf,labelsep=space]{caption}
\usepackage{float}
\usepackage{fancyhdr}
\usepackage{fnpos}
\usepackage[english]{babel}
\addto{\captionsenglish}{}
\usepackage{array}
\usepackage{droidsans}
\usepackage{charter}
\usepackage[T1]{fontenc}
\usepackage[usenames,dvipsnames]{xcolor}
\usepackage{setspace}
\usepackage[compact]{titlesec}
\usepackage{hyperref}

\usepackage{amsmath}
\usepackage{amssymb}
\usepackage{algorithm}
\usepackage{algpseudocode}

\definecolor{cream}{RGB}{222,217,201}

\begin{document}

\pagestyle{fancy}
\thispagestyle{plain}
\fancypagestyle{plain}{
\renewcommand{\headrulewidth}{0pt}}

\makeFNbottom
\makeatletter
\renewcommand\LARGE{\@setfontsize\LARGE{15pt}{17}}
\renewcommand\Large{\@setfontsize\Large{12pt}{14}}
\renewcommand\large{\@setfontsize\large{10pt}{12}}
\renewcommand\footnotesize{\@setfontsize\footnotesize{7pt}{10}}
\makeatother

\renewcommand{\thefootnote}{\fnsymbol{footnote}}
\renewcommand\footnoterule{\vspace*{1pt}%
\color{cream}\hrule width 3.5in height 0.4pt \color{black}\vspace*{5pt}} 
\setcounter{secnumdepth}{5}

\makeatletter 
\renewcommand\@biblabel[1]{#1}            
\renewcommand\@makefntext[1]%
{\noindent\makebox[0pt][r]{\@thefnmark\,}#1}
\makeatother 
\renewcommand{\figurename}{\small{Fig.}~}
\sectionfont{\sffamily\Large}
\subsectionfont{\normalsize}
\subsubsectionfont{\bf}
\setstretch{1.125} 
\setlength{\skip\footins}{0.8cm}
\setlength{\footnotesep}{0.25cm}
\setlength{\jot}{10pt}
\titlespacing*{\section}{0pt}{4pt}{4pt}
\titlespacing*{\subsection}{0pt}{15pt}{1pt}

\fancyfoot{}
\fancyfoot[LO,RE]{\vspace{-7.1pt}\includegraphics[height=9pt]{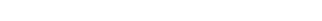}}
\fancyfoot[CO]{\vspace{-7.1pt}\hspace{13.2cm}\includegraphics{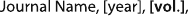}}
\fancyfoot[CE]{\vspace{-7.2pt}\hspace{-14.2cm}\includegraphics{head_foot/RF}}
\fancyfoot[RO]{\footnotesize{\sffamily{1--\pageref{LastPage} ~\textbar  \hspace{2pt}\thepage}}}
\fancyfoot[LE]{\footnotesize{\sffamily{\thepage~\textbar\hspace{3.45cm} 1--\pageref{LastPage}}}}
\fancyhead{}
\renewcommand{\headrulewidth}{0pt} 
\renewcommand{\footrulewidth}{0pt}
\setlength{\arrayrulewidth}{1pt}
\setlength{\columnsep}{6.5mm}
\setlength\bibsep{1pt}
\raggedbottom

\makeatletter 
\newlength{\figrulesep} 
\setlength{\figrulesep}{0.5\textfloatsep} 

\newcommand{\topfigrule}{\vspace*{-1pt}%
\noindent{\color{cream}\rule[-\figrulesep]{\columnwidth}{1.5pt}} }

\newcommand{\botfigrule}{\vspace*{-2pt}%
\noindent{\color{cream}\rule[\figrulesep]{\columnwidth}{1.5pt}} }

\newcommand{\dblfigrule}{\vspace*{-1pt}%
\noindent{\color{cream}\rule[-\figrulesep]{\textwidth}{1.5pt}} }

\makeatother

\twocolumn[
  \begin{@twocolumnfalse}
{\includegraphics[height=30pt]{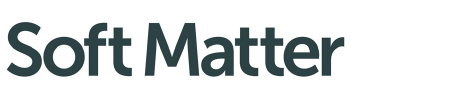}\hfill\raisebox{0pt}[0pt][0pt]{\includegraphics[height=55pt]{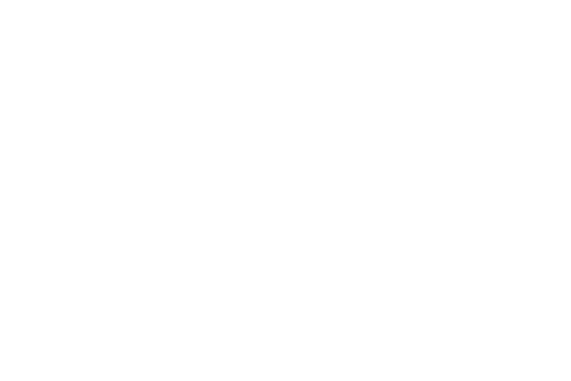}}\\[1ex]
\includegraphics[width=18.5cm]{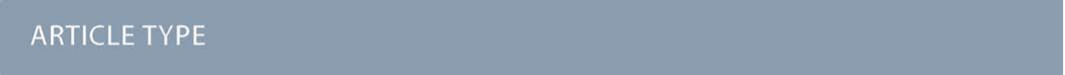}}\par
\vspace{1em}
\sffamily
\begin{tabular}{m{4.5cm} p{13.5cm} }

\includegraphics{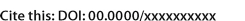} & \noindent\LARGE{\textbf{Computational Self-Assembly of a Six-Fold\newline Chiral Quasicrystal}} \\
\vspace{0.3cm} & \vspace{0.3cm} \\

& \noindent\large{Nydia Roxana Varela-Rosales\textit{$^{a, b}$}, Michael Engel\textit{$^a$}} \\

\includegraphics{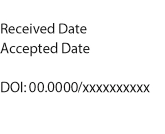} & \noindent\normalsize{
    Quasicrystals are unique materials characterized by long-range order without periodicity. They are observed in systems such as metallic alloys, soft matter, and particle simulations. Unlike periodic crystals, which are invariant under real-space symmetry operations, quasicrystals possess symmetry that requires description by a space group in reciprocal space. In this study, we report the self-assembly of a six-fold chiral quasicrystal using molecular dynamics simulations of a two-dimensional particle system. The particles interact via the Lennard-Jones-Gauss pair potential and are subjected to a periodic substrate potential. We confirm the presence of chiral symmetry through diffraction patterns and order parameters, revealing unique local motifs in both real and reciprocal space. The quasicrystal's properties, including the tiling structure and symmetry and the extent of diffuse scattering, are strongly influenced by substrate potential depth and temperature. Our results provide insights into the mechanisms of chiral quasicrystal formation and the role and potential of external fields in tailoring quasicrystal structures.
} \\
\end{tabular}
  \end{@twocolumnfalse} \vspace{0.6cm}
]

\renewcommand*\rmdefault{bch}\normalfont\upshape
\rmfamily
\section*{}
\vspace{-1cm}

\footnotetext{\textit{$^{a}$~Institute for Multiscale Simulation, IZNF, Friedrich-Alexander-Universität Erlangen-Nürnberg, 91058 Erlangen, Germany. E-mail: michael.engel@fau.de}}

\footnotetext{\textit{$^{b}$~MathAM-OIL, AIST, c/o Advanced Institute for Materials Research, Tohoku University, 980-8577 Sendai, Japan. E-mail: nydia.roxana.varela.rosales.e2@tohoku.ac.jp}}


\section{Introduction}

A periodic crystal is considered chiral if it cannot be superimposed onto its mirror image by a proper isometry. A proper isometry is a distance-preserving transformation that maintains handedness, and it can be expressed as a combination of a translation and a rotation. The space group of a chiral crystal includes only proper isometries. Notably, only 65 out of the 230 three-dimensional space groups and 5 out of the 17 two-dimensional space groups, also known as wallpaper groups, exhibit this property. Although chiral crystals are relatively rare, they possess unique properties that make them valuable in various applications,~\cite{riehl2011} such as the rotation of polarized light in optical devices, the efficacy and safety of enantiomer-specific drugs in pharmaceuticals, and the facilitation of second harmonic generation in nonlinear optics for advanced laser technologies.

Chirality becomes even more intriguing in aperiodic crystals and tilings. A notable example is the hat monotile, also known as `Einstein', a chiral 13-sided convex polygon. Tiling the plane with the hat tile produces an aperiodic tiling that is distinguishable from its mirror image.~\cite{Smith2024, Socolar2023} Another example is the surface of certain families of viruses,~\cite{Konevtsova2012} which can be described by a finite tiling on the surface of a sphere created from a quasicrystal through the action of a chiral phason strain field. Discussing chirality in quasicrystals requires extending classical crystallographic concepts, as quasicrystals do not repeat and lack a space group defined in real space. Instead, the Fourier transform of a quasicrystal, with its dense set of sharp peaks, reveals regular features that can be used to generalize the concept of symmetry,~\cite{Rokhsar_1988, Lifshitz_1996} associating quasicrystals with a space group in reciprocal space.~\cite{Janot1993, Walter2009} This allows for the consideration of chiral quasicrystals as those quasicrystals with a chiral space group. While chirality in aperiodic tilings has been considered for some time in the mathematics community~\cite{Fujita2009,Baake2013, Baake2017}, we are not aware of chiral quasicrystals reported to form spontaneously in either experimental or simulation studies.

\begin{figure*}
    \centering
    \includegraphics[width=1\linewidth]{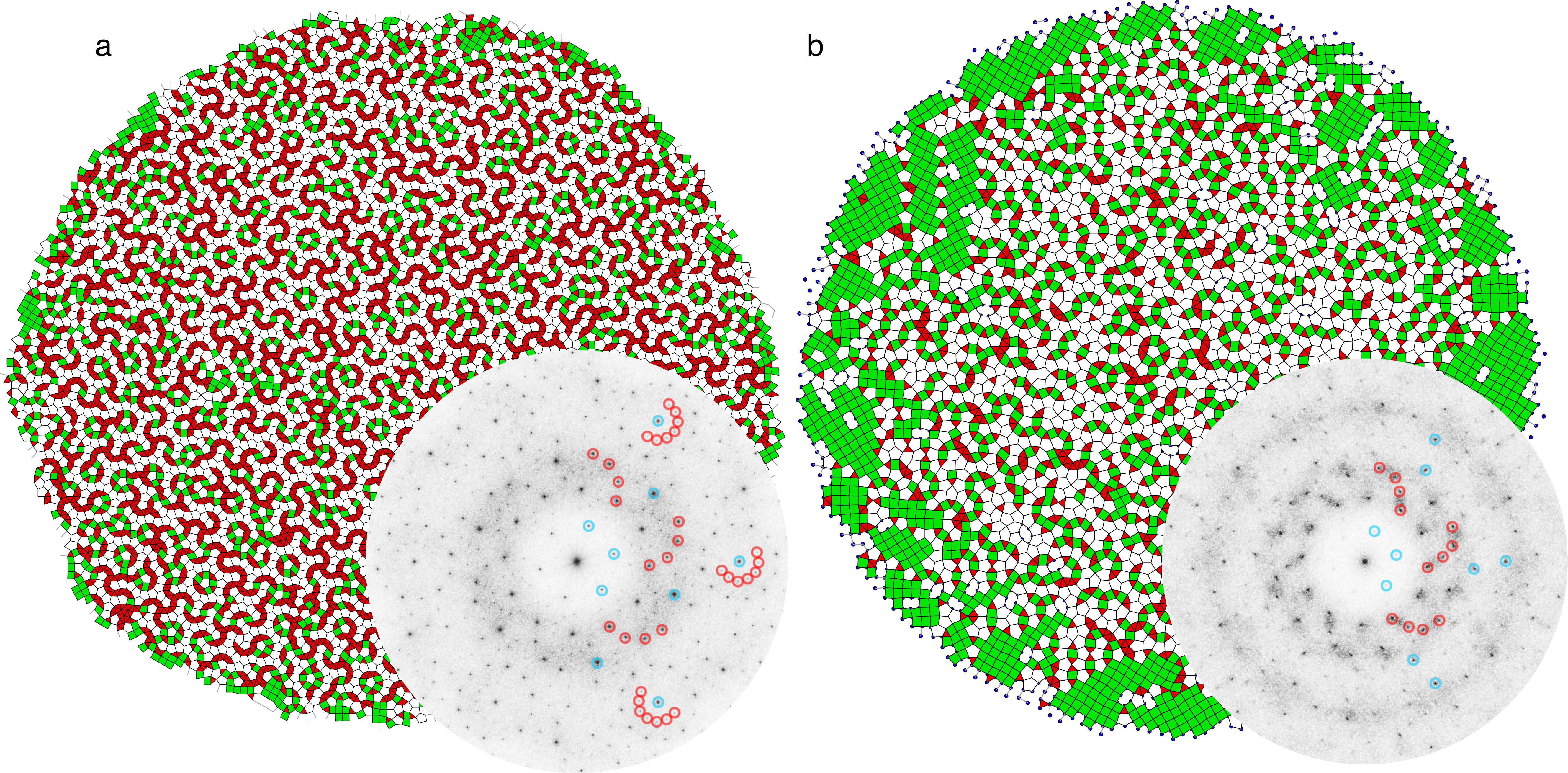}
    \caption{MD simulation snapshots of the CQC at the parameter sets (a)~$(T,\varepsilon)=(0.25,0.5)$ and (b)~$(T,\varepsilon)=(0.3,0.4)$. In these snapshots, nearest-neighbor bonds create a tiling pattern, where triangle tiles are colored red and square tiles are colored green. The insets show diffraction patterns, time-averaged to reduce phonon noise and with chiral symmetry and six-fold symmetry features highlighted using small red and blue circles, respectively. In snapshot (b), the solid-gas interface is wetted by square tiles and diffuse scattering is a more pronounced compared to (a).}
    \label{fig:patterns_of_DiffPattern_diff_epsilon_values_two_cqc}
\end{figure*}

In our search for a chiral quasicrystal, we utilize a two-dimensional model system of particles interacting via the Lennard-Jones-Gauss potential, which is known to form multiple achiral quasicrystals.~\cite{Engel_2007,Engel_F-L_validation} Similar achiral two-dimensional~\cite{Widom1987, Leung_dod-QC_1989, Dzugutov_1993, Dod_Roth1998, Quandt_1999, Engel_2007, Engel_2010, Marjolein_2012,3D_dod_universal_2017, Dotera2014, Dotera2019} and three-dimensional~\cite{Roth1995,Engel2015,Noya2021,Noya2021dod} quasicrystals have commonly been self-assembled in simulations involving particle mixtures of varying sizes or multiple wells in the interaction potential. To induce chirality, we introduce an additional periodic substrate potential. In a previous study~\cite{Varela2023}, we examined the effects of weak substrate potentials on the stability of various approximants of the dodecagonal quasicrystal and other incommensurately modulated phases, as indicated by satellite peaks in the diffraction patterns. In this study, we demonstrate that significantly stronger substrate potentials, compared to our previous work, cause the achiral quasicrystal to spontaneously transform into a chiral quasicrystal (Fig.~\ref{fig:patterns_of_DiffPattern_diff_epsilon_values_two_cqc}).

The concept of using substrates to design new quasicrystals has been explored for some time. Thin films can be induced into quasiperiodic order when grown on substrates with quasiperiodic symmetry.~\cite{McGrath_templated_QC_mol_ord_2014,experimental_ex_LEED,experimental_ex_nanoepitaxy, Weisskopf_Y_sim_sub_deca_2007} Another approach involves templating quasicrystals by tuning the interference patterns of laser beams.~\cite{arch_tiles_part2_Schmiedeberg2010, arch_tiles_Mikhael, arch_tiles_Mikhael2008, Schmiedeberg_2D_quasicrystSubst, Jagannathan2014_eight_fold_laser} Strong laser beam intensity can force particles to adhere to the symmetry of the laser field, whereas weak fields allow particles to retain their native, periodic configuration. By adjusting the laser field strength, one can interpolate between these cases,~\cite{Lowen_structural_trapped_2013} creating a competition between the symmetry of the template and the native symmetry that the particles prefer in the absence of the template. This method can also target approximants (periodic crystals approximating the structure of a quasicrystal) by using substrates to favor specific tiles.~\cite{monolayer_Patrykiejew2009,sq_sub_Schmiedeberg_2013}

\section{Methods} \label{Methods}

\subsection{Molecular dynamics simulations}

We performed two-dimensional molecular dynamics (MD) simulations using the HOOMD-blue simulation package~\cite{hoomd_2015,hoomd_2008}. The particle systems were modeled with the Lennard-Jones-Gauss (LJG) potential
\begin{equation}
V_\text{LJG}(r) = \frac{1}{r^{12}} -  \frac{2}{r^{6}} - {{\varepsilon}_\text{LJG}} \exp \left( \frac{-(r - r_0)^2 }{ 2\sigma^{2} }  \right).
\label{eq:LJG}
\end{equation}
The parameters were chosen as $\varepsilon_\text{LJG}=1.8$, ${r_0}=1.42$, and $\sigma^{2}=0.042$ based on our previous study~\cite{Engel_F-L_validation} where a dodecagonal quasicrystal was observed. Simulations were conducted in the $NVT$ ensemble using a Nosé–Hoover thermostat and effectively open boundary conditions on the simulated crystal. The Boltzmann constant was set to 1. To mimic the effect of a substrate with hexagonal lattice symmetry, we applied an external potential,
\begin{equation}\label{eq:ext_pot}
V_\text{ext} = \frac{\varepsilon\varepsilon_\text{LJG}}{6T} \sum_{i=0}^{2}\cos(2\pi\mathbf{k}_{i} \cdot \mathbf{r}/\lambda)
\end{equation}
with $\mathbf{k}_{i} =(\cos(\pi i/3),\sin(\pi i/3))$. The parameter $\varepsilon$ represents the substrate potential depth, and $\lambda$ is the substrate periodicity. To maintain consistency with our previous work and to align the assembly diagram in Fig.~2 with that in Fig.~5 of Ref.~\cite{Varela2023}, we set the substrate periodicity to $\lambda=0.5$ and measure substrate potential depth relative to kinetic energy, i.e.\ use the prefactor $\varepsilon/T$. This choice ensures that the substrate has constant influence on particle dynamics relative to kinetic energy.

Particles were initially positioned within a central circle at a number density of 0.3. Over time, the systems crystallized into circular clusters exhibiting solid-gas coexistence. This setup was chosen to ensure open boundaries around the crystalline cluster. To construct an assembly diagram across varying $\varepsilon$ and $T$ parameter values, we performed extensive MD simulations, each comprising $10^8$ MD integration steps. This duration was sufficient for the potential energy to reach a plateau, allowing us to accurately determine phase boundaries. These boundaries serve as a guide to understanding the system's phase behavior.

\subsection{Diffraction symmetry order parameters} \label{Sec:DiffractionOP}

Diffraction patterns are crucial for characterizing quasicrystals. In this section we discuss their significance and methodology in detail. For a given system of particles, we construct a continuous particle density $\rho(\mathbf{r})$ by either time-averaging multiple snapshots or by applying a narrow Gaussian convolution to smear the particle positions in a single snapshot. The Fourier transform of this density is computed as $\rho(\mathbf{k})=\int_B\exp(i\mathbf{k}\cdot \mathbf{r})\rho(\mathbf{r})\,d\mathbf{r}$, where the integration is performed over the simulation box $B$. From this transformation, we obtain the structure factor or diffraction pattern, denoted as $S(\mathbf{k})=\|\rho(\mathbf{k})\|^2$. Diffraction patterns are generated using the INJAVIS visualization package~\cite{Engel2021injavis}, with parameters set to a zoom factor of 0.5, an image size of 1024 bins, and a peak width of 5.

To analyze symmetry in quasicrystal patterns, we calculate order parameters in reciprocal space. The similarity between two diffraction patterns, $S_1$ and $S_2$, is quantified using the Pearson correlation coefficient
\begin{equation}
    \text{Corr}(S_1, S_2)=\frac{\text{Cov}(S_1, S_2)}{\sqrt{\text{Cov}(S_1, S_1)\text{Cov}(S_2, S_2)}},
\end{equation}
where the covariance is defined as
\begin{equation}
    \text{Cov}(S_1, S_2)=\int S_1(\mathbf{k})S_2(\mathbf{k})\,d\mathbf{k}.
\end{equation}

To evaluate rotational symmetry, we use the $n$-fold rotation matrix
\begin{equation}
    R_n = \begin{pmatrix}
        \cos(\frac{2\pi}{n}) & -\sin(\frac{2\pi}{n}) \\
        \sin(\frac{2\pi}{n}) &  \cos(\frac{2\pi}{n})
    \end{pmatrix}.
\end{equation}
Perfect $n$-fold rotational symmetry requires the diffraction pattern to remain invariant under this rotation, expressed mathematically as $S\left((R_n)^m\mathbf{k}\right)= S\left(\mathbf{k}\right)$ for all integers $m$ and wave vectors $\mathbf{k}$. This can be concisely stated as $S\circ R_n^m \equiv S$. By definition, diffraction patterns are always invariant under two-fold rotation, i. e., $S\circ R_2 \equiv S$. We define order parameters for six-fold and twelve-fold rotational symmetry by averaging over certain equivalent orientations as
\begin{align} 
    r_{6}  &= \text{Corr}(S\circ R_{6}  + S\circ R_{6} ^2, S),\\
    r_{12} &= \text{Corr}(S\circ R_{12} + S\circ R_{12}^5, S).
\end{align}
For the six-fold symmetry order parameter $r_6$, we avoid trivial correlations due to the exact relationship $S\circ R_{2n}^{m} \equiv S\circ R_{2n}^{n+m}$. Similarly, for the twelve-fold symmetry order parameter $r_{12}$, we exclude correlations caused by the presence of lower rotational symmetries.

Quantifying chiral symmetry involves additional steps. We use the mirror operation matrix 
\begin{equation}
    M_\theta = \begin{pmatrix}
        -\cos(2\theta) & \sin(2\theta) \\
         \sin(2\theta) & \cos(2\theta)
    \end{pmatrix}
\end{equation}
where the mirror axis is rotated by an angle $\theta$ relative to the $x$-axis. Because the orientation of the mirror axis is unknown, we define the chiral order parameter $\chi$ as the solution to an optimization problem,
\begin{equation}
    \chi= 1 - \frac{\max_{\theta\in[0,\pi)}\text{Corr}(S\circ M_\theta,S)- r_6}{1-r_6},
\end{equation}
which is solved numerically. The subtraction of $r_6$ and normalization enhances the detection of chirality in two-dimensional structures with six-fold symmetry. We propose that similar strategies can be applied to quantify chirality in other particle structures with n-fold symmetry by appropriately modifying the term $r_6$.

\subsection{Neural network classification}

We employed a convolutional neural network (CNN) to classify regions within the $\varepsilon$-$T$ parameter space where the chiral quasicrystal is present. This CNN is designed for binary classification, using diffraction patterns of the chiral quasicrystal as training data. The model architecture comprises five convolutional layers with increasing complexity, featuring filter sizes ranging from 32 to 512. Each convolutional layer is followed by a 2x2 max pooling layer to reduce spatial dimensions and highlight important features. Following the convolutional layers, the model includes a dense network segment consisting of four fully connected layers with 1024, 512, 256, and 128 neurons, respectively. To prevent overfitting a dropout layer with a rate of 0.2 is included. The network concludes with a sigmoid activation layer that outputs the probability of the input being classified as part of the chiral quasicrystal class. The Python code implementing this CNN architecture is available in the Zenodo repository accompanying this work.

\section{Results}   \label{Results}

\subsection{Discovery of a chiral quasicrystal}

\begin{figure*}[ht!]
    \centering
    \includegraphics[width=1\linewidth]{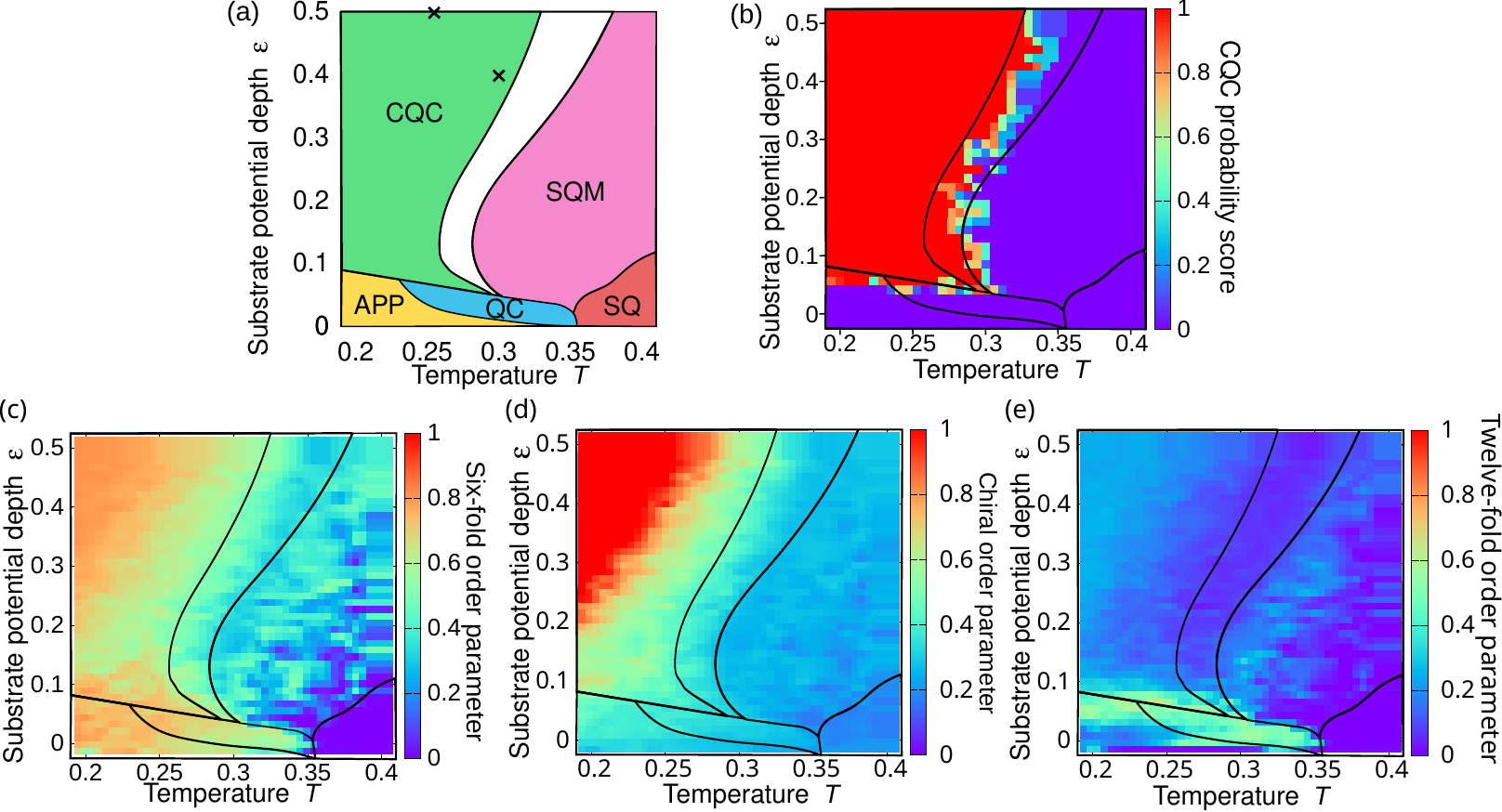}
    \caption{(a)~Assembly diagram constructed from simulation snapshots. The chiral quasicrystal (CQC) phase is observed at low temperatures and high substrate potential depths. Other identified phases include the quasicrystal (QC), approximants (APP), square phase (SQ), and square modulations (SQM). The white region indicates the coexistence of CQC and SQM phases. Two cross-like symbols mark the positions in the stability diagram corresponding to the snapshots shown in Fig.~\ref{fig:patterns_of_DiffPattern_diff_epsilon_values_two_cqc}. (b)~CQC probability score obtained from the CNN model. Order parameters quantify the presence of different symmetries: (c)~Six-fold order is prominent in the QC, CQC, and APP regions. (d)~Chiral order is particularly high in the region where the CQC with low diffuse scattering is found. (e)~Twelve-fold order is significant in the QC region.}
    \label{fig:stabilityDiagram}
\end{figure*}

We investigated substrate potential depths in the range of $0\leq\varepsilon\leq 0.5$, which is significantly broader than previously studied ranges.~\cite{Varela2023} Numerous MD simulations were conducted, and the final snapshots were analyzed through visual inspections of tilings and examination of diffraction patterns. The observed phase behavior is summarized in an assembly diagram (Fig.~\ref{fig:stabilityDiagram}a). This diagram extends Fig.~5 of Ref.~\cite{Varela2023} to significantly higher $\varepsilon$ values, confirming the presence of the chiral quasicrystal (CQC). It also includes the previously reported square phase (SQ) and its modulated version (SQM) at higher temperatures, as well as the achiral dodecagonal quasicrystal (QC) and various approximants (APP).

\begin{figure}[ht!]
    \centering
    \includegraphics[width=1\linewidth]{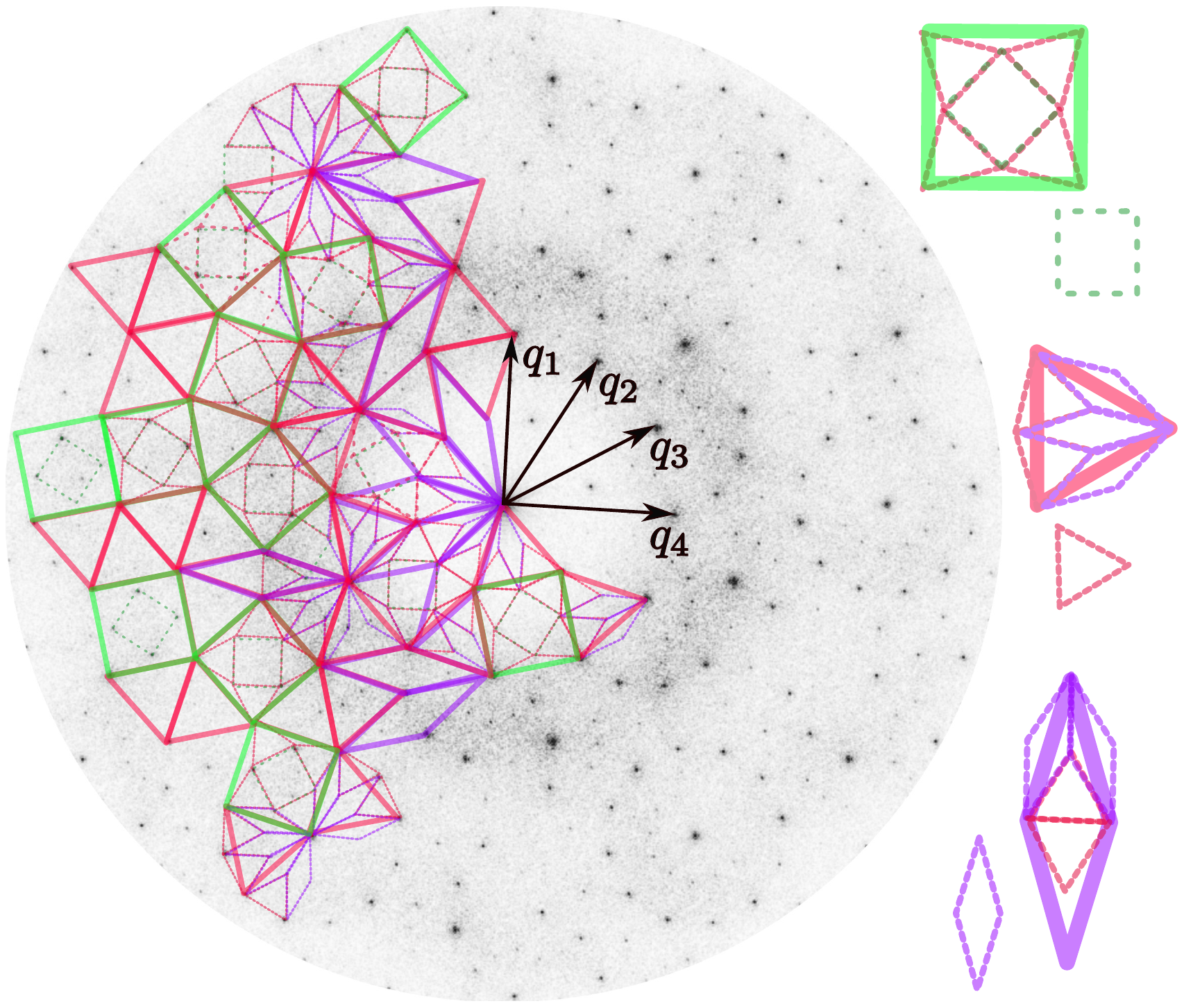}
    \caption{Analysis of peak positions in the diffraction pattern of Fig.~\ref{fig:patterns_of_DiffPattern_diff_epsilon_values_two_cqc}(a). $q_i$ are four basis vectors spanning the reciprocal lattice. Square, triangle, and rhombus tiles are overlaid onto the diffraction pattern. Each tile can undergo further substitution (right side) with irrational inflation factor $(1+\sqrt{3})/\sqrt{2}$. This demonstrates that each diffraction peak position can be written as an integer multiple of the four basis vectors.}
    \label{fig:six-fold_QC_diffraction}
\end{figure}

Two snapshots from the CQC region are shown in Fig.~\ref{fig:patterns_of_DiffPattern_diff_epsilon_values_two_cqc}. The first snapshot (Fig.~\ref{fig:patterns_of_DiffPattern_diff_epsilon_values_two_cqc}a) was taken deep within the CQC region at $T=0.2$ and $\varepsilon=0.5$. Nearest-neighbor bonds form a tiling composed of triangle tiles (red), square tiles (green), and pentagon tiles (white). This self-assembled tiling exhibits long-range order but lacks periodicity. The colored tiles interconnect in a meandering manner, creating a spiral-like network pattern. While the tiling does not exhibit a clear inflation symmetry in real space, the diffraction pattern reveals distinct peaks along with weak diffuse scattering. The set of diffraction peaks can be indexed by four basis vectors (Fig.~\ref{fig:six-fold_QC_diffraction}) and displays six-fold symmetry without mirror symmetry (Fig.~\ref{fig:patterns_of_DiffPattern_diff_epsilon_values_two_cqc}(a)). The absence of mirror symmetry indicates chiral symmetry in reciprocal space. Together, these observations are sufficient to identify the pattern as a chiral quasicrystal. The second snapshot (Fig.~\ref{fig:patterns_of_DiffPattern_diff_epsilon_values_two_cqc}b) was taken closer to the stability boundary of the CQC with SQM at higher $T$ and lower $\varepsilon$. Here, the solid-gas interface of the self-assembled tiling becomes populated with patches of square tiles. Simultaneously, the number of triangle tiles decreases within the CQC, suppressing the meandering network responsible for the chiral nature of the CQC. Because of a higher number of structural defects in the second snapshots, the diffraction pattern exhibits more pronounced diffuse scattering.

We delineate the stability region of the CQC using a CNN model trained to recognize CQC patterns (Fig.~\ref{fig:stabilityDiagram}b). The CNN analysis confirms that the CQC spans a significant portion of the assembly diagram at low temperatures ($T\lesssim 0.3$) and high substrate potential depth ($\varepsilon\gtrsim 0.1$). Clearly, the CQC forms robustly over a broad parameter range. Furthermore, the CQC is not a unique crystallographic structure specified by a single tiling. As we move through the parameter range, the tile composition changes continuously (visible also in Fig.~\ref{fig:patterns_of_DiffPattern_diff_epsilon_values_two_cqc}). This adaptability allows the CQC to respond flexibly to thermodynamic constraints ($T$) and the constraints imposed by the external potential ($\varepsilon$).

\begin{figure*}[ht!]
    \centering
    \includegraphics[width=1\linewidth]{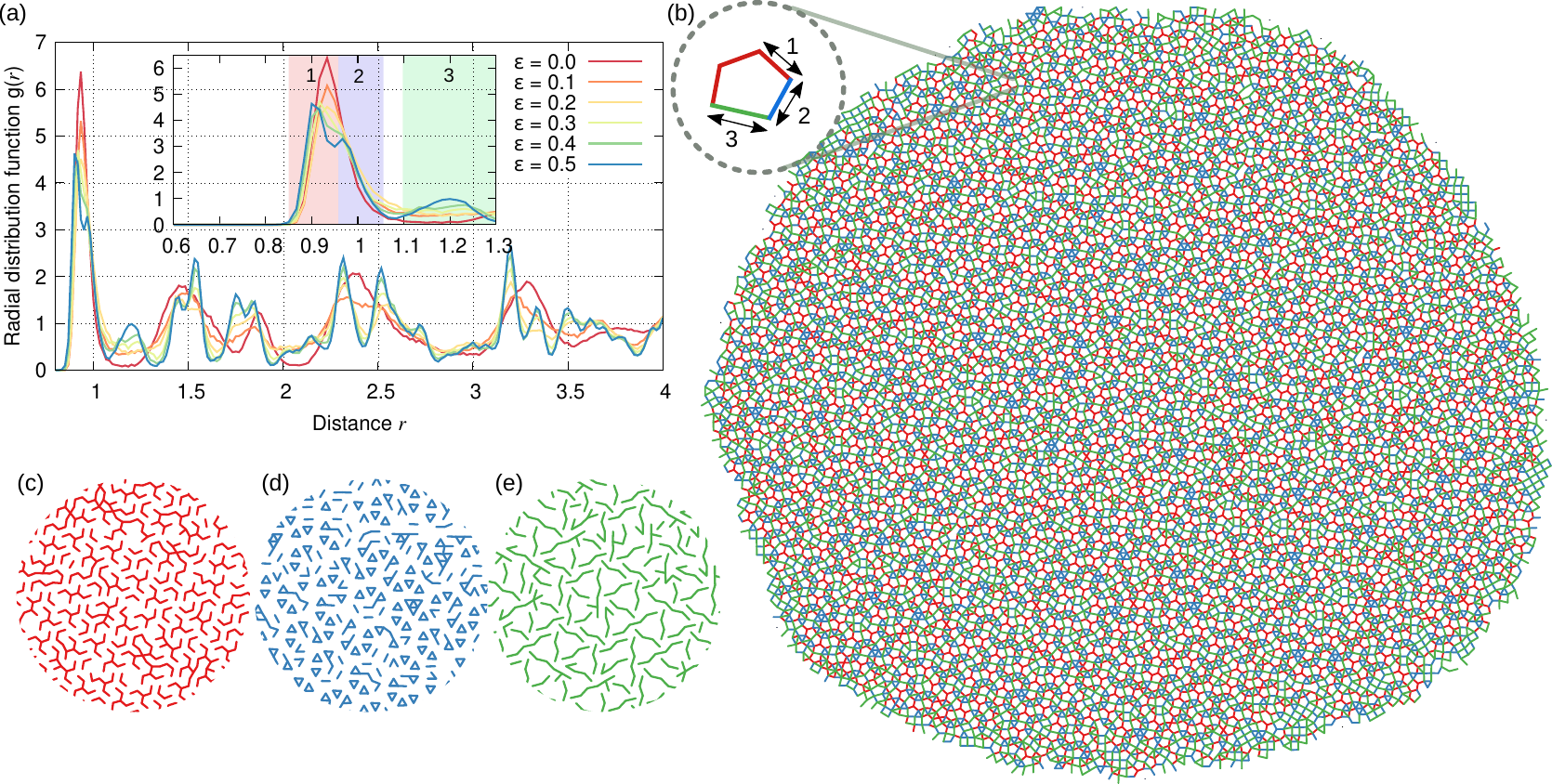}
    \caption{(a)~RDFs averaged from MD trajectories at various $\varepsilon$ values with $T=0.2$. The inset provides a detailed view of the short-distance range. Shaded areas indicate the cutoffs used to construct bond networks, highlighting short nearest-neighbor bonds (red), long nearest-neighbor bonds (blue), and secondary peak bonds (green). (b)~A simulation snapshot at $\varepsilon=0.5$, illustrating the three types of bonds identified in the RDF. In pentagon tiles, there are typically three red bonds, one green bond, and one blue bond. (c-e)~The same snapshot as in (b), with each panel displaying only one type of bond: (c) short nearest-neighbor bonds (red), (d) long nearest-neighbor bonds (blue), and (e) secondary peak bonds (green).}
    \label{fig:RDF_diff_epsilon_values}
\end{figure*}

Finally, we apply the diffraction symmetry order parameters introduced in Sec.~\ref{Sec:DiffractionOP} to quantify the presence of crystallographic symmetries. The six-fold symmetry order parameter (Fig.~\ref{fig:stabilityDiagram}c) effectively identifies six-fold symmetry in the CQC, QC, and APP regions. All approximants in this system exhibit six-fold symmetry.~\cite{Varela2023} Six-fold order decreases toward the coexistence region with SQM due to the formation of surface patches with square tiles, as illustrated in Fig.~\ref{fig:patterns_of_DiffPattern_diff_epsilon_values_two_cqc}b. Similarly, the chiral order parameter (Fig.~\ref{fig:stabilityDiagram}d) shows a strong correlation with the six-fold order parameter in the CQC region and follows a similar trend. Other phases in the assembly diagram do not exhibit chirality. Lastly, the twelve-fold order parameter (Fig.~\ref{fig:stabilityDiagram}e) shows significant values only in the QC region, with the APP and CQC regions lacking notable twelve-fold symmetry.

\subsection{Bond network in the CQC}

To better understand the intricate structure of the CQC tiling, we analyze the radial distribution function (RDF). The RDF represents the average particle density in a ring around a reference particle and changes significantly with the substrate potential depth $\varepsilon$ (Fig.~\ref{fig:RDF_diff_epsilon_values}a). As we transition from the APP regime ($\varepsilon<0.1$) to the CQC regime ($\varepsilon>0.1$), the first peak in the range $0.85\leq r\leq 1.05$ splits into what we term short nearest-neighbor bonds (red shaded range in the inset) and long nearest-neighbor bonds (blue shaded range). Additionally, a secondary nearest-neighbor peak emerges at $r\approx 1.2$ (green shaded range). We call all bonds associated with these three ranges as nearest neighbors, because they form the backbone of the tiling pattern. The changes in the RDF with $\varepsilon$ indicate a more diverse local order and thus a higher structural complexity of the CQC compared to the APP. Similar splittings of the first peak have been reported in other quasicrystal-forming simulations.~\cite{Mizuguchi2009, Ryltsev2015} The observation that the changes in the RDF are continuous throughout the CQC regime confirms that the CQC structure is not a single unique tiling but can change continuously across the parameter regime in the assembly diagram where it is the preferred structure. 

To interpret the significance of the first RDF peak splitting and the secondary peak, we colored the bonds in a simulation snapshot according to the three shading colors in the RDF. The resulting colored tiling is shown in Fig.~\ref{fig:RDF_diff_epsilon_values}b. The figure demonstrates that triangles and squares have various types of bonds, often of mixed type in a single tile. Pentagons most commonly have three red bonds, one blue bond, and one green bond. This implies that the tiles are slightly deformed in the CQC. It also shows how the CQC can adapt to the geometric constraints induced by the periodic substrate by rearranging locally. The availability of multiple competing local environments increases the possibilities for the CQC to respond to these constraints.

Plotting the bond networks for the three colors separately (Figs.~\ref{fig:RDF_diff_epsilon_values}c-e), provides additional insights. The bond network reveals larger-scale structures, which may hint at weak inflation symmetry and some regular order on a larger scale. It remains unclear whether the tiling can eventually transform into new types of chiral approximants at low temperatures or form a more ordered chiral quasicrystal than the random tiling with high degree of randomness we typically observe in our simulations. We also observe chirality more clearly in the bond networks of individual colors. The red bond network displays three-legged motifs resembling a triskelion, with counter-clockwise triskelions dominating in Fig.~\ref{fig:RDF_diff_epsilon_values}c. The green bond network features wavy lines forming open triangular cages. These cages have three narrow openings consistently arranged counter-clockwise near the tips, thus exhibiting chiral symmetry (Fig.~\ref{fig:RDF_diff_epsilon_values}e).

\subsection{Variation of substrate periodicity}

\begin{figure*}[ht!]
    \centering
    \includegraphics[width=0.7\linewidth]{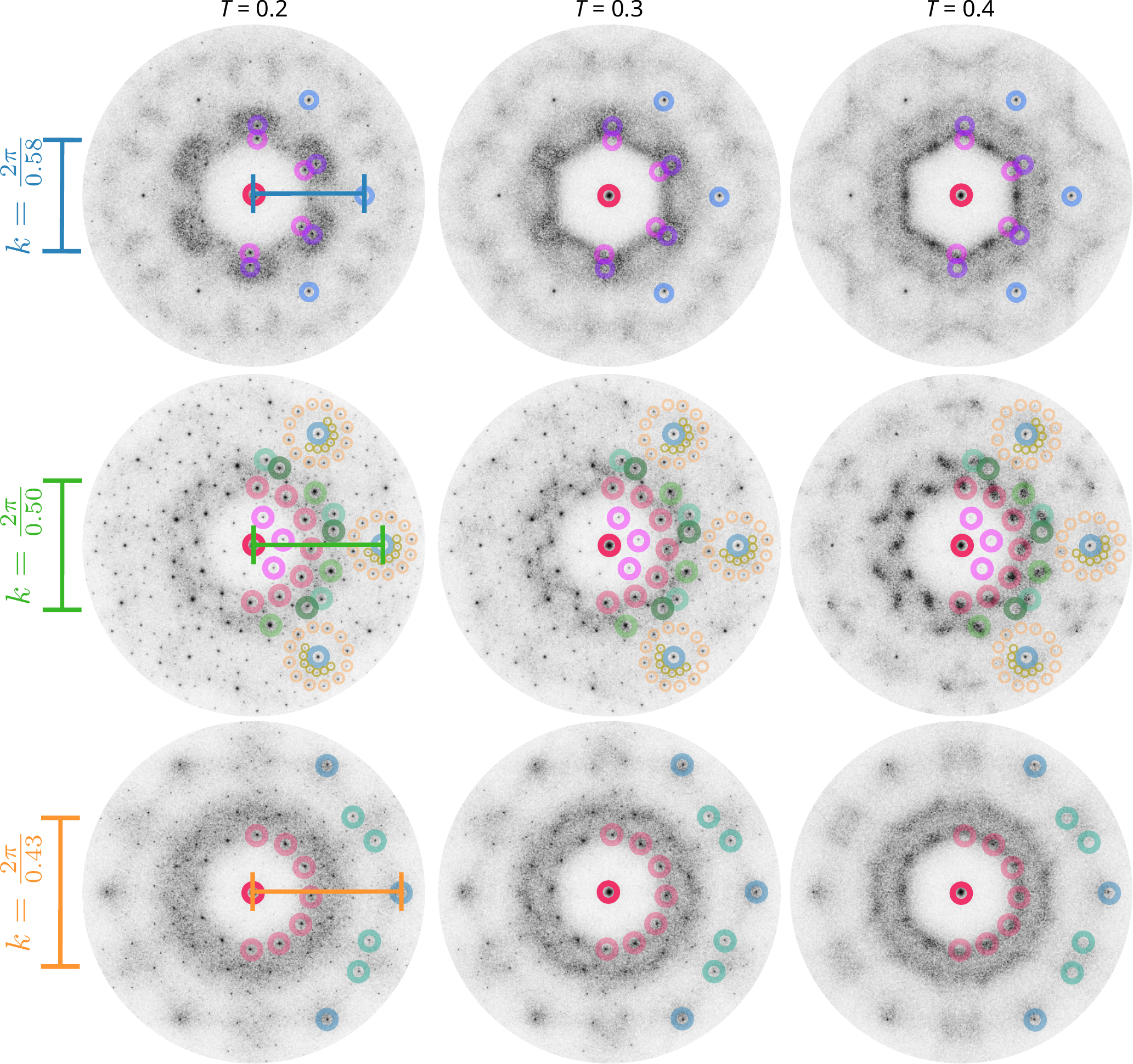}
    \caption{Diffraction patterns of simulation snapshots for three substrate periodicities $\lambda\in[0.43, 0.5, 0.58]$ (vertical axis) and three temperatures $T\in[0.2,0.3,0.4]$ (horizontal axis), at fixed substrate potential depth $\varepsilon=0.5$. Each $\lambda$ value corresponds to a wave vector $k=2\pi/\lambda$, as indicated by the colored scale bars. The diffraction patterns reveal characteristic peaks at these wave vectors, underscoring the significant role of substrate periodicity in shaping the system's structure. Additional notable diffraction peaks are highlighted with colored circles. Chirality is observed only at $\lambda = 0.5$, whereas other $\lambda$ values lead to significant diffuse scattering.}
    \label{fig:DiffPattern_diff_period_values}
\end{figure*}

In this section, we explore how the substrate periodicity parameter $\lambda$ influences the stabilization of the CQC. We compare simulations using the primary value of $\lambda=0.5$ from this study with simulations at higher and lower values, $\lambda=0.58$ and $\lambda=0.43$, respectively. Our analysis of the diffraction patterns reveals a clear correlation between $\lambda$ and specific diffraction peaks (Fig.~\ref{fig:DiffPattern_diff_period_values}). The diffraction patterns are highly sensitive to changes in substrate periodicity. Sharp diffraction spots appear only at $\lambda=0.5$, whereas the other $\lambda$ values produce patterns with significant diffuse scattering, indicating structural disorder. This suggests that $\lambda = 0.5$, a parameter initially chosen somewhat accidentally, albeit inspired by an experimental system,~\cite{Varela2023} is nearly optimal for the emergence of the CQC. Deviations from $\lambda=0.5$, whether higher or lower, result in the disappearance of the CQC. Future research should investigate how $\lambda = 0.5$ effectively shifts the system away from the QC phase and promotes six-fold chiral symmetry. Understanding this mechanism will further elucidate the factors contributing to CQC stabilization.

\section{Conclusions}

This study demonstrated the self-assembly of a six-fold chiral quasicrystal through molecular dynamics simulations of a two-dimensional particle system. The particles interacted via the Lennard-Jones-Gauss pair potential and were influenced by a periodic hexagonal substrate potential. Remarkably, they formed a chiral quasicrystal despite the substrate’s achiral nature. This unexpected formation underscored the complex interplay between particle interactions and external fields in quasicrystal formation causing a spontaneous breaking of mirror symmetry.

We confirmed the presence of chiral symmetry using distinct diffraction patterns, unique local motifs, and a chiral order parameter, which provided evidence of local chirality in real space and global chirality in reciprocal space. Our analysis of the radial distribution function revealed the emergence of multiple nearest-neighbor bond lengths as the substrate potential depth increased, indicating enhanced structural complexity and diversity in local ordering. Additionally, the quasicrystal was stable only at an optimal substrate periodicity.

Chiral quasicrystal formation remains a relatively unexplored area in quasicrystal research. While our simulations provided valuable insights into the self-assembly processes and symmetry properties of chiral quasicrystals, further investigation is needed to elucidate the underlying mechanisms driving their formation and to determine the optimal geometric descriptions of their tiling patterns. Future research should focus on exploring these mechanisms and the potential applications of chiral quasicrystals in materials science.

\section*{Conflicts of interest}
There are no conflicts to declare.

\section*{Acknowledgements}
We acknowledge funding by Deutsche Forschungsgemeinschaft (DFG) through grant EN 905/4-1. HPC resources provided by the Erlangen National High Performance Computing Center (NHR@FAU) under the NHR project b168dc are gratefully acknowledged. NHR funding is provided by federal and Bavarian state authorities. NHR@FAU hardware is partially funded by DFG under Project-ID 440719683.

\section*{Data Availability Statement}

Data for this article, including images/graphics and Python scripts are available at Zenodo at \url{https://doi.org/10.5281/zenodo.13208339}.

\section*{Table of Content entry}

\includegraphics[width=8cm]{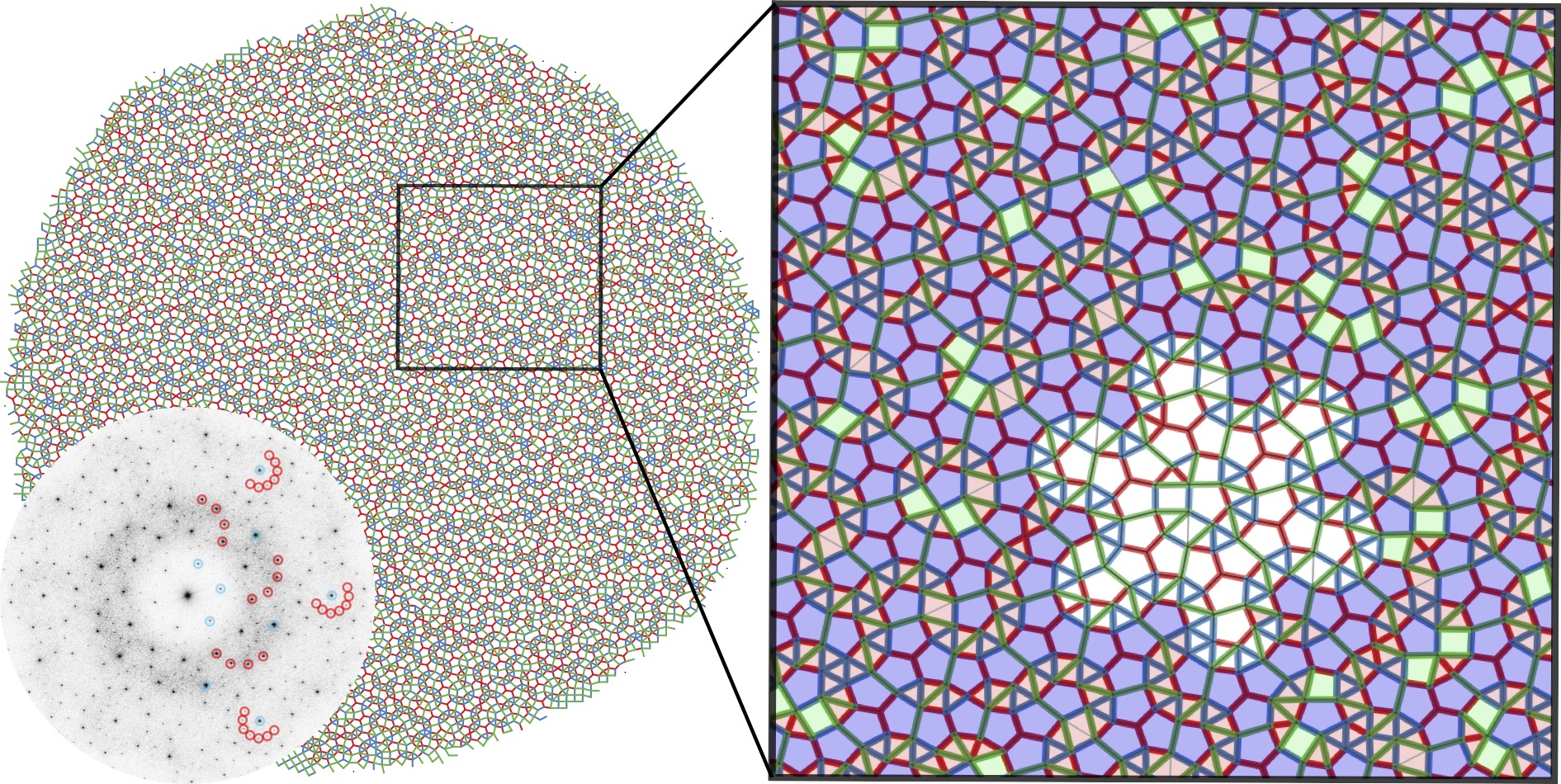}\\
Six-fold chiral quasicrystal with diffraction pattern and a close-up of a six-fold star-shaped pattern.

\balance

\bibliography{ChiralQuasicrystal} 

\providecommand*{\mcitethebibliography}{\thebibliography}
\csname @ifundefined\endcsname{endmcitethebibliography}
{\let\endmcitethebibliography\endthebibliography}{}
\begin{mcitethebibliography}{45}
\providecommand*{\natexlab}[1]{#1}
\providecommand*{\mciteSetBstSublistMode}[1]{}
\providecommand*{\mciteSetBstMaxWidthForm}[2]{}
\providecommand*{\mciteBstWouldAddEndPuncttrue}
  {\def\EndOfBibitem{\unskip.}}
\providecommand*{\mciteBstWouldAddEndPunctfalse}
  {\let\EndOfBibitem\relax}
\providecommand*{\mciteSetBstMidEndSepPunct}[3]{}
\providecommand*{\mciteSetBstSublistLabelBeginEnd}[3]{}
\providecommand*{\EndOfBibitem}{}
\mciteSetBstSublistMode{f}
\mciteSetBstMaxWidthForm{subitem}
{(\emph{\alph{mcitesubitemcount}})}
\mciteSetBstSublistLabelBeginEnd{\mcitemaxwidthsubitemform\space}
{\relax}{\relax}

\bibitem[Riehl(2011)]{riehl2011}
J.~Riehl, \emph{Mirror-Image Asymmetry: An Introduction to the Origin and Consequences of Chirality}, Wiley, 2011\relax
\mciteBstWouldAddEndPuncttrue
\mciteSetBstMidEndSepPunct{\mcitedefaultmidpunct}
{\mcitedefaultendpunct}{\mcitedefaultseppunct}\relax
\EndOfBibitem
\bibitem[Smith \emph{et~al.}(2024)Smith, Myers, Kaplan, and Goodman-Strauss]{Smith2024}
D.~Smith, J.~S. Myers, C.~S. Kaplan and C.~Goodman-Strauss, \emph{Combinatorial Theory}, 2024, \textbf{4}, 1--91\relax
\mciteBstWouldAddEndPuncttrue
\mciteSetBstMidEndSepPunct{\mcitedefaultmidpunct}
{\mcitedefaultendpunct}{\mcitedefaultseppunct}\relax
\EndOfBibitem
\bibitem[Socolar(2023)]{Socolar2023}
J.~E.~S. Socolar, \emph{Phys. Rev. B}, 2023, \textbf{108}, 224109\relax
\mciteBstWouldAddEndPuncttrue
\mciteSetBstMidEndSepPunct{\mcitedefaultmidpunct}
{\mcitedefaultendpunct}{\mcitedefaultseppunct}\relax
\EndOfBibitem
\bibitem[Konevtsova \emph{et~al.}(2012)Konevtsova, Rochal, and Lorman]{Konevtsova2012}
O.~V. Konevtsova, S.~B. Rochal and V.~L. Lorman, \emph{Phys. Rev. Lett.}, 2012, \textbf{108}, 038102\relax
\mciteBstWouldAddEndPuncttrue
\mciteSetBstMidEndSepPunct{\mcitedefaultmidpunct}
{\mcitedefaultendpunct}{\mcitedefaultseppunct}\relax
\EndOfBibitem
\bibitem[Rokhsar \emph{et~al.}(1988)Rokhsar, Wright, and Mermin]{Rokhsar_1988}
D.~S. Rokhsar, D.~C. Wright and N.~D. Mermin, \emph{Acta Cryst.}, 1988, \textbf{44}, 197--211\relax
\mciteBstWouldAddEndPuncttrue
\mciteSetBstMidEndSepPunct{\mcitedefaultmidpunct}
{\mcitedefaultendpunct}{\mcitedefaultseppunct}\relax
\EndOfBibitem
\bibitem[Lifshitz(1996)]{Lifshitz_1996}
R.~Lifshitz, \emph{Physica A}, 1996, \textbf{232}, 633--647\relax
\mciteBstWouldAddEndPuncttrue
\mciteSetBstMidEndSepPunct{\mcitedefaultmidpunct}
{\mcitedefaultendpunct}{\mcitedefaultseppunct}\relax
\EndOfBibitem
\bibitem[Janot(1993)]{Janot1993}
C.~Janot, \emph{Prof. R. Soc. A}, 1993, \textbf{442}, 113--127\relax
\mciteBstWouldAddEndPuncttrue
\mciteSetBstMidEndSepPunct{\mcitedefaultmidpunct}
{\mcitedefaultendpunct}{\mcitedefaultseppunct}\relax
\EndOfBibitem
\bibitem[Walter and Deloudi(2009)]{Walter2009}
S.~Walter and S.~Deloudi, \emph{Crystallography of Quasicrystals: Concepts, Methods and Structures}, Springer Berlin Heidelberg, 2009\relax
\mciteBstWouldAddEndPuncttrue
\mciteSetBstMidEndSepPunct{\mcitedefaultmidpunct}
{\mcitedefaultendpunct}{\mcitedefaultseppunct}\relax
\EndOfBibitem
\bibitem[Fujita(2009)]{Fujita2009}
N.~Fujita, \emph{Acta Cryst. A}, 2009, \textbf{A65}, 342--351\relax
\mciteBstWouldAddEndPuncttrue
\mciteSetBstMidEndSepPunct{\mcitedefaultmidpunct}
{\mcitedefaultendpunct}{\mcitedefaultseppunct}\relax
\EndOfBibitem
\bibitem[Baake and Grimm(2013)]{Baake2013}
M.~Baake and U.~Grimm, \emph{Aperiodic Order Vol. 1, sec. 6.5.1}, Cambridge University Press, 2013\relax
\mciteBstWouldAddEndPuncttrue
\mciteSetBstMidEndSepPunct{\mcitedefaultmidpunct}
{\mcitedefaultendpunct}{\mcitedefaultseppunct}\relax
\EndOfBibitem
\bibitem[Baake and Grimm(2017)]{Baake2017}
M.~Baake and U.~Grimm, \emph{Aperiodic Order Vol. 2, p. 31}, Cambridge University Press, 2017\relax
\mciteBstWouldAddEndPuncttrue
\mciteSetBstMidEndSepPunct{\mcitedefaultmidpunct}
{\mcitedefaultendpunct}{\mcitedefaultseppunct}\relax
\EndOfBibitem
\bibitem[Engel and Trebin(2007)]{Engel_2007}
M.~Engel and H.-R. Trebin, \emph{Phys. Rev. Lett.}, 2007, \textbf{98}, 225505\relax
\mciteBstWouldAddEndPuncttrue
\mciteSetBstMidEndSepPunct{\mcitedefaultmidpunct}
{\mcitedefaultendpunct}{\mcitedefaultseppunct}\relax
\EndOfBibitem
\bibitem[Engel(2011)]{Engel_F-L_validation}
M.~Engel, \emph{Phys. Rev. Lett.}, 2011, \textbf{106}, 095504\relax
\mciteBstWouldAddEndPuncttrue
\mciteSetBstMidEndSepPunct{\mcitedefaultmidpunct}
{\mcitedefaultendpunct}{\mcitedefaultseppunct}\relax
\EndOfBibitem
\bibitem[Widom \emph{et~al.}(1987)Widom, Strandburg, and Swendsen]{Widom1987}
M.~Widom, K.~J. Strandburg and R.~H. Swendsen, \emph{Phys. Rev. Lett.}, 1987, \textbf{58}, 706--709\relax
\mciteBstWouldAddEndPuncttrue
\mciteSetBstMidEndSepPunct{\mcitedefaultmidpunct}
{\mcitedefaultendpunct}{\mcitedefaultseppunct}\relax
\EndOfBibitem
\bibitem[Leung \emph{et~al.}(1989)Leung, Henley, and Chester]{Leung_dod-QC_1989}
P.~W. Leung, C.~L. Henley and G.~V. Chester, \emph{Phys. Rev. B}, 1989, \textbf{39}, 446--458\relax
\mciteBstWouldAddEndPuncttrue
\mciteSetBstMidEndSepPunct{\mcitedefaultmidpunct}
{\mcitedefaultendpunct}{\mcitedefaultseppunct}\relax
\EndOfBibitem
\bibitem[Dzugutov(1993)]{Dzugutov_1993}
M.~Dzugutov, \emph{Phys. Rev. Lett.}, 1993, \textbf{70}, 2924--2927\relax
\mciteBstWouldAddEndPuncttrue
\mciteSetBstMidEndSepPunct{\mcitedefaultmidpunct}
{\mcitedefaultendpunct}{\mcitedefaultseppunct}\relax
\EndOfBibitem
\bibitem[Roth and G{\"a}hler(1998)]{Dod_Roth1998}
J.~Roth and F.~G{\"a}hler, \emph{Eur. Phys. J. B}, 1998, \textbf{6}, 425--445\relax
\mciteBstWouldAddEndPuncttrue
\mciteSetBstMidEndSepPunct{\mcitedefaultmidpunct}
{\mcitedefaultendpunct}{\mcitedefaultseppunct}\relax
\EndOfBibitem
\bibitem[Quandt and Teter(1999)]{Quandt_1999}
A.~Quandt and M.~P. Teter, \emph{Phys. Rev. B}, 1999, \textbf{59}, 8586--8592\relax
\mciteBstWouldAddEndPuncttrue
\mciteSetBstMidEndSepPunct{\mcitedefaultmidpunct}
{\mcitedefaultendpunct}{\mcitedefaultseppunct}\relax
\EndOfBibitem
\bibitem[Engel \emph{et~al.}(2010)Engel, Umezaki, Trebin, and Odagaki]{Engel_2010}
M.~Engel, M.~Umezaki, H.-R. Trebin and T.~Odagaki, \emph{Phys. Rev. B}, 2010, \textbf{82}, 134206\relax
\mciteBstWouldAddEndPuncttrue
\mciteSetBstMidEndSepPunct{\mcitedefaultmidpunct}
{\mcitedefaultendpunct}{\mcitedefaultseppunct}\relax
\EndOfBibitem
\bibitem[van~der Linden \emph{et~al.}(2012)van~der Linden, Doye, and Louis]{Marjolein_2012}
M.~N. van~der Linden, J.~P.~K. Doye and A.~A. Louis, \emph{J. Chem. Phys.}, 2012, \textbf{136}, 054904\relax
\mciteBstWouldAddEndPuncttrue
\mciteSetBstMidEndSepPunct{\mcitedefaultmidpunct}
{\mcitedefaultendpunct}{\mcitedefaultseppunct}\relax
\EndOfBibitem
\bibitem[Ryltsev and Chtchelkatchev(2017)]{3D_dod_universal_2017}
R.~Ryltsev and N.~Chtchelkatchev, \emph{Soft Matter}, 2017, \textbf{13}, 5076--5082\relax
\mciteBstWouldAddEndPuncttrue
\mciteSetBstMidEndSepPunct{\mcitedefaultmidpunct}
{\mcitedefaultendpunct}{\mcitedefaultseppunct}\relax
\EndOfBibitem
\bibitem[Dotera \emph{et~al.}(2014)Dotera, Oshiro, and Ziherl]{Dotera2014}
T.~Dotera, T.~Oshiro and P.~Ziherl, \emph{Nature}, 2014, \textbf{506}, 208--211\relax
\mciteBstWouldAddEndPuncttrue
\mciteSetBstMidEndSepPunct{\mcitedefaultmidpunct}
{\mcitedefaultendpunct}{\mcitedefaultseppunct}\relax
\EndOfBibitem
\bibitem[Nakakura \emph{et~al.}(2019)Nakakura, Ziherl, Matsuzawa, and Dotera]{Dotera2019}
J.~Nakakura, P.~Ziherl, J.~Matsuzawa and T.~Dotera, \emph{Nat. Commun.}, 2019, \textbf{10}, 4235\relax
\mciteBstWouldAddEndPuncttrue
\mciteSetBstMidEndSepPunct{\mcitedefaultmidpunct}
{\mcitedefaultendpunct}{\mcitedefaultseppunct}\relax
\EndOfBibitem
\bibitem[Roth \emph{et~al.}(1995)Roth, Schilling, and Trebin]{Roth1995}
J.~W. Roth, R.~Schilling and H.-R. Trebin, \emph{Phys. Rev. B}, 1995, \textbf{51}, 15833--15840\relax
\mciteBstWouldAddEndPuncttrue
\mciteSetBstMidEndSepPunct{\mcitedefaultmidpunct}
{\mcitedefaultendpunct}{\mcitedefaultseppunct}\relax
\EndOfBibitem
\bibitem[Engel \emph{et~al.}(2015)Engel, Damasceno, Phillips, and Glotzer]{Engel2015}
M.~Engel, P.~F. Damasceno, C.~L. Phillips and S.~C. Glotzer, \emph{Nat. Mater.}, 2015, \textbf{14}, 109--116\relax
\mciteBstWouldAddEndPuncttrue
\mciteSetBstMidEndSepPunct{\mcitedefaultmidpunct}
{\mcitedefaultendpunct}{\mcitedefaultseppunct}\relax
\EndOfBibitem
\bibitem[Noya \emph{et~al.}(2021)Noya, Wong, Llombart, and Doye]{Noya2021}
E.~G. Noya, C.~K. Wong, P.~Llombart and J.~P.~K. Doye, \emph{Nature}, 2021, \textbf{596}, 367--371\relax
\mciteBstWouldAddEndPuncttrue
\mciteSetBstMidEndSepPunct{\mcitedefaultmidpunct}
{\mcitedefaultendpunct}{\mcitedefaultseppunct}\relax
\EndOfBibitem
\bibitem[Tracey \emph{et~al.}(2021)Tracey, Noya, and Doye]{Noya2021dod}
D.~F. Tracey, E.~G. Noya and J.~P.~K. Doye, \emph{J. Chem. Phys.}, 2021, \textbf{154}, 194505\relax
\mciteBstWouldAddEndPuncttrue
\mciteSetBstMidEndSepPunct{\mcitedefaultmidpunct}
{\mcitedefaultendpunct}{\mcitedefaultseppunct}\relax
\EndOfBibitem
\bibitem[Varela-Rosales and Engel(2024)]{Varela2023}
N.~R. Varela-Rosales and M.~Engel, \emph{Soft Matter}, 2024, \textbf{20}, 2915--2925\relax
\mciteBstWouldAddEndPuncttrue
\mciteSetBstMidEndSepPunct{\mcitedefaultmidpunct}
{\mcitedefaultendpunct}{\mcitedefaultseppunct}\relax
\EndOfBibitem
\bibitem[Smerdon \emph{et~al.}(2014)Smerdon, Young, Lowe, Hars, Yadav, Hesp, Dhanak, Tsai, Sharma, and McGrath]{McGrath_templated_QC_mol_ord_2014}
J.~A. Smerdon, K.~M. Young, M.~Lowe, S.~S. Hars, T.~P. Yadav, D.~Hesp, V.~R. Dhanak, A.~P. Tsai, H.~R. Sharma and R.~McGrath, \emph{Nano Lett.}, 2014, \textbf{14}, 1184--1189\relax
\mciteBstWouldAddEndPuncttrue
\mciteSetBstMidEndSepPunct{\mcitedefaultmidpunct}
{\mcitedefaultendpunct}{\mcitedefaultseppunct}\relax
\EndOfBibitem
\bibitem[Ferralis \emph{et~al.}(2004)Ferralis, Diehl, Pussi, Lindroos, Fisher, and Jenks]{experimental_ex_LEED}
N.~Ferralis, R.~D. Diehl, K.~Pussi, M.~Lindroos, I.~Fisher and C.~J. Jenks, \emph{Phys. Rev. B}, 2004, \textbf{69}, 075410\relax
\mciteBstWouldAddEndPuncttrue
\mciteSetBstMidEndSepPunct{\mcitedefaultmidpunct}
{\mcitedefaultendpunct}{\mcitedefaultseppunct}\relax
\EndOfBibitem
\bibitem[Flückiger \emph{et~al.}(2003)Flückiger, Weisskopf, Erbudak, Lüscher, and Kortan]{experimental_ex_nanoepitaxy}
T.~Flückiger, Y.~Weisskopf, M.~Erbudak, R.~Lüscher and A.~R. Kortan, \emph{Nano Lett.}, 2003, \textbf{3}, 1717--1721\relax
\mciteBstWouldAddEndPuncttrue
\mciteSetBstMidEndSepPunct{\mcitedefaultmidpunct}
{\mcitedefaultendpunct}{\mcitedefaultseppunct}\relax
\EndOfBibitem
\bibitem[Bilki \emph{et~al.}(2007)Bilki, Erbudak, Mungan, and Weisskopf]{Weisskopf_Y_sim_sub_deca_2007}
B.~Bilki, M.~Erbudak, M.~Mungan and Y.~Weisskopf, \emph{Phys. Rev. B}, 2007, \textbf{75}, 045437\relax
\mciteBstWouldAddEndPuncttrue
\mciteSetBstMidEndSepPunct{\mcitedefaultmidpunct}
{\mcitedefaultendpunct}{\mcitedefaultseppunct}\relax
\EndOfBibitem
\bibitem[Schmiedeberg \emph{et~al.}(2010)Schmiedeberg, Mikhael, Rausch, Roth, Helden, Bechinger, and Stark]{arch_tiles_part2_Schmiedeberg2010}
M.~Schmiedeberg, J.~Mikhael, S.~Rausch, J.~Roth, L.~Helden, C.~Bechinger and H.~Stark, \emph{Eur. Phys. J. E}, 2010, \textbf{32}, 25--34\relax
\mciteBstWouldAddEndPuncttrue
\mciteSetBstMidEndSepPunct{\mcitedefaultmidpunct}
{\mcitedefaultendpunct}{\mcitedefaultseppunct}\relax
\EndOfBibitem
\bibitem[Mikhael \emph{et~al.}(2011)Mikhael, Gera, Bohlein, and Bechinger]{arch_tiles_Mikhael}
J.~Mikhael, G.~Gera, T.~Bohlein and C.~Bechinger, \emph{Soft Matter}, 2011, \textbf{7}, 1352--1357\relax
\mciteBstWouldAddEndPuncttrue
\mciteSetBstMidEndSepPunct{\mcitedefaultmidpunct}
{\mcitedefaultendpunct}{\mcitedefaultseppunct}\relax
\EndOfBibitem
\bibitem[Mikhael \emph{et~al.}(2008)Mikhael, Roth, Helden, and Bechinger]{arch_tiles_Mikhael2008}
J.~Mikhael, J.~Roth, L.~Helden and C.~Bechinger, \emph{Nature}, 2008, \textbf{454}, 501--504\relax
\mciteBstWouldAddEndPuncttrue
\mciteSetBstMidEndSepPunct{\mcitedefaultmidpunct}
{\mcitedefaultendpunct}{\mcitedefaultseppunct}\relax
\EndOfBibitem
\bibitem[Schmiedeberg and Stark(2008)]{Schmiedeberg_2D_quasicrystSubst}
M.~Schmiedeberg and H.~Stark, \emph{Phys. Rev. Lett.}, 2008, \textbf{101}, 218302\relax
\mciteBstWouldAddEndPuncttrue
\mciteSetBstMidEndSepPunct{\mcitedefaultmidpunct}
{\mcitedefaultendpunct}{\mcitedefaultseppunct}\relax
\EndOfBibitem
\bibitem[Jagannathan and Duneau(2014)]{Jagannathan2014_eight_fold_laser}
A.~Jagannathan and M.~Duneau, \emph{Eur. Phys. J. B}, 2014, \textbf{87}, 149\relax
\mciteBstWouldAddEndPuncttrue
\mciteSetBstMidEndSepPunct{\mcitedefaultmidpunct}
{\mcitedefaultendpunct}{\mcitedefaultseppunct}\relax
\EndOfBibitem
\bibitem[Costa~Campos \emph{et~al.}(2013)Costa~Campos, Apolinario, and L\"owen]{Lowen_structural_trapped_2013}
L.~Q. Costa~Campos, S.~W.~S. Apolinario and H.~L\"owen, \emph{Phys. Rev. E}, 2013, \textbf{88}, 042313\relax
\mciteBstWouldAddEndPuncttrue
\mciteSetBstMidEndSepPunct{\mcitedefaultmidpunct}
{\mcitedefaultendpunct}{\mcitedefaultseppunct}\relax
\EndOfBibitem
\bibitem[Patrykiejew \emph{et~al.}(2009)Patrykiejew, R{\.{z}}ysko, and Soko{\l}owski]{monolayer_Patrykiejew2009}
A.~Patrykiejew, W.~R{\.{z}}ysko and S.~Soko{\l}owski, \emph{Adsorption}, 2009, \textbf{15}, 254--263\relax
\mciteBstWouldAddEndPuncttrue
\mciteSetBstMidEndSepPunct{\mcitedefaultmidpunct}
{\mcitedefaultendpunct}{\mcitedefaultseppunct}\relax
\EndOfBibitem
\bibitem[Neuhaus \emph{et~al.}(2013)Neuhaus, Marechal, Schmiedeberg, and L\"owen]{sq_sub_Schmiedeberg_2013}
T.~Neuhaus, M.~Marechal, M.~Schmiedeberg and H.~L\"owen, \emph{Phys. Rev. Lett.}, 2013, \textbf{110}, 118301\relax
\mciteBstWouldAddEndPuncttrue
\mciteSetBstMidEndSepPunct{\mcitedefaultmidpunct}
{\mcitedefaultendpunct}{\mcitedefaultseppunct}\relax
\EndOfBibitem
\bibitem[Glaser \emph{et~al.}(2015)Glaser, Nguyen, Anderson, Lui, Spiga, Millan, Morse, and Glotzer]{hoomd_2015}
J.~Glaser, T.~D. Nguyen, J.~A. Anderson, P.~Lui, F.~Spiga, J.~A. Millan, D.~C. Morse and S.~C. Glotzer, \emph{Comput. Phys. Commun.}, 2015, \textbf{192}, 97--107\relax
\mciteBstWouldAddEndPuncttrue
\mciteSetBstMidEndSepPunct{\mcitedefaultmidpunct}
{\mcitedefaultendpunct}{\mcitedefaultseppunct}\relax
\EndOfBibitem
\bibitem[Anderson \emph{et~al.}(2008)Anderson, Lorenz, and Travesset]{hoomd_2008}
J.~A. Anderson, C.~D. Lorenz and A.~Travesset, \emph{J. Comput. Phys.}, 2008, \textbf{227}, 5342--5359\relax
\mciteBstWouldAddEndPuncttrue
\mciteSetBstMidEndSepPunct{\mcitedefaultmidpunct}
{\mcitedefaultendpunct}{\mcitedefaultseppunct}\relax
\EndOfBibitem
\bibitem[Engel(2021)]{Engel2021injavis}
M.~Engel, \emph{INJAVIS — INteractive JAva VISualization}, 2021, \url{https://doi.org/10.5281/zenodo.4639570}\relax
\mciteBstWouldAddEndPuncttrue
\mciteSetBstMidEndSepPunct{\mcitedefaultmidpunct}
{\mcitedefaultendpunct}{\mcitedefaultseppunct}\relax
\EndOfBibitem
\bibitem[Mizuguchi and Odagaki(2009)]{Mizuguchi2009}
T.~Mizuguchi and T.~Odagaki, \emph{Phys. Rev. E}, 2009, \textbf{79}, 051501\relax
\mciteBstWouldAddEndPuncttrue
\mciteSetBstMidEndSepPunct{\mcitedefaultmidpunct}
{\mcitedefaultendpunct}{\mcitedefaultseppunct}\relax
\EndOfBibitem
\bibitem[Ryltsev \emph{et~al.}(2015)Ryltsev, Klumov, and Chtchelkatchev]{Ryltsev2015}
R.~Ryltsev, B.~Klumov and N.~Chtchelkatchev, \emph{Soft Matter}, 2015, \textbf{11}, 6991--6998\relax
\mciteBstWouldAddEndPuncttrue
\mciteSetBstMidEndSepPunct{\mcitedefaultmidpunct}
{\mcitedefaultendpunct}{\mcitedefaultseppunct}\relax
\EndOfBibitem
\end{mcitethebibliography}
\bibliographystyle{rsc} 

\end{document}